Working PaperWorking Paper

# Navigating the AI-Energy Nexus with Geopolitical Insight

Nidhi Kalra, Robin Wang, and Ismael Arciniegas Rueda

RAND Global and Emerging Risks

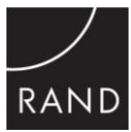

# About This Working Paper

As of early 2025, the U.S. is the clear global leader in Artificial Intelligence (AI), but this leadership is at risk. Advancing AI will require, among other things, significant amounts of new, reliable, sustainable electricity. Where this energy is supplied will influence where, globally, future AI is developed, owned, and used. Much research is underway to understand how the AI-energy nexus could play out.

In the U.S. and Western democracies, markets play a central role in resource allocation and economic development. However, key competitors such as China and the Gulf Countries have far more centralized and authoritarian decision making. In these countries, the state has strong influence over a much broader aspect of the economy and society and can overcome or even counter prevailing market forces.

In this Working Paper, we discuss the relative capacities for U.S. competitors to allocate resources for strategic objectives like AI energy supremacy. We suggest that understanding competitiveness requires using geopolitically informed data and analysis, and that there are key market and non-market options to increase U.S. competitiveness, borrowing strengths from both domestic and foreign political economies.

For many who study national policy in China or the gulf countries, the message in this Working Paper will be familiar. It will also be of interest to technologists, researchers, and policymakers addressing general U.S. AI competitiveness. The Working Paper aims to inform policymakers on the U.S. capabilities in providing energy for AI. It will aim to inform those less versed in the political economies of U.S. competitors and their implications for America's global AI leadership.

## *Technology and Security Policy Center*

RAND Global and Emerging Risks is a division of RAND that delivers rigorous and objective public policy research on the most consequential challenges to civilization and global security. This work was undertaken by the division's Technology and Security Policy Center, which explores how high-consequence, dual-use technologies change the global competition and threat environment, then develops policy and technology options to advance the security of the United States, its allies and partners, and the world. For more information, contact tasp@rand.org.

## *Funding*

Founders Pledge, Charlottes och Fredriks Stiftelse, Good Ventures, Jaan Tallinn, Longview, Open Philanthropy, and Waking Up Foundation. A complete list of donors and funders is available at www.rand.org/TASP. RAND donors and grantors have no influence over research findings or recommendations.

*Acknowledgments*

We thank our colleagues Joel Predd, Bryan Boling, and Hye Min Park for their helpful suggestions. We appreciate help from Gregory Smith in shepherding this paper along.

*About the Authors*

Nidhi Kalra is a senior information scientist at the RAND Corporation. Her research focuses on autonomous vehicle policy, climate change adaptation, and tools and methods that help people and organizations make better decisions amid deep uncertainty. She holds a Ph.D. in robotics.

Robin Wang is a Graduate student at the RAND School of Public Policy. His research at RAND focuses on climate adaptation, infrastructure resilience, and developing modelling tools to support policy decisions to address cascading risks. Robin holds a MPA degree from the London School of Economics and Political Science.

Ismael Arciniegas Rueda is a Senior Economist at RAND where he currently works on several energy related topics such as China energy challenge, Artificial Intelligence in power grids, electric utilities resilience to disasters and energy markets. He holds a Ph.D in Economics.

# Contents

## Introduction

As of early 2025, the U.S. has remained the clear global leader in AI (Maslej et al., 2024), but this leadership is at risk. Sustaining a competitive edge in AI requires meeting the growing demands for hardware, human talent, land, energy, water, and other resources, amidst competing sectoral needs. And, while frontier model development remains concentrated in the U.S. and China, recent breakthroughs like Claude 3 and DeepSeek-V2 demonstrate a shift toward greater model efficiency, composability, and accessibility (Singh, 2025). These shifts may enable a wider range of countries and firms to participate in AI deployment and innovation, even without frontier-scale compute infrastructure. The strategic geography of AI competitiveness is likely to broaden—not only based on who builds the models, but also on who adapts, governs, and uses them effectively.

Energy supply is particularly critical, as advancing AI will require significant amounts of new, reliable, sustainable electricity (Fist and Datta, 2024). Where this energy is supplied will influence where, globally, future AI and Artificial General Intelligence (AGI) is developed, owned, and controlled (Metz and Mickle, 2024).

To maintain its lead, the U.S. may need to significantly increase its net electricity generation capacity and the scale and reliability of its transmission infrastructure. Yet, as Fist and Datta (2024) show, the U.S. lags key AI competitors, notably China. Since 2010, China and Brazil have on average added 50 gigawatt (GW) and 15GW of new capacity each year, respectively, while the U.S. has added only about 1 GW per year. In some of those years, total capacity in the U.S. declined rather than increased.

Embedded in this difference are differences in speed, scale, and quality of new power supply. In the U.S., regulatory and legal processes – to establish load certainty, engage stakeholders, and authorize procurement of new power – govern resource and transmission expansion and may take many years. Resource operations (interconnecting, permitting, construction, etc.) can begin after these decisions, and can take even more time if new transmission lines are required to deliver power.

Other countries also have leading capabilities in key energy technologies. China is by far the lead in deploying new nuclear power stations: from 2013 to 2023, China added 37 nuclear reactors, twice as many as the rest of the world combined ("China is building nuclear reactors faster than any other country," 2023). However, many U.S. AI companies are looking to nuclear energy to power the next generation of models (Aldrete et al., 2024).

At the same time, China and other competitors in the AI leadership race have different power supply challenges. Electricity supply has been less reliable in China, for example, hampering its economic, environmental, and other ambitions (Qiu et al., 2024). China has also struggled to balance its rapid growth in installed renewable energy capacity and its slower growth in transmission, leading to high curtailment rates — when available renewable electricity is wasted because the grid cannot absorb or transmit it. (Cheng et al., 2023).

The energy demands of AI will also not be uniform across AI competitor countries; they will vary based on factors such as the size and purposes of AI ecosystem and infrastructure a country pursues, the climatic and other conditions that will drive cooling needs for data centers, and commitments to reducing local air pollution, greenhouse gas (GHG) emissions, or other environmental goals.

The U.S.'s sustained AI leadership may depend on its ability to address these energy demands effectively to outpace rivals better positioned to meet them. Alternatively, other countries may be better positioned to meet these energy demands, accelerating their domestic AI efforts and overtake U.S. technology leadership or attract U.S. technology companies to relocate their training, inference, and research and development activities.

## Understanding Risks to U.S. AI Leadership Requires Examining Competitors' Energy Sectors Through Market and Non-Market Lenses

A few studies have explored the geopolitical landscape of energy and AI development. The Institute for Progress has released a series of analyses – Compute in America – which goes into the technical requirements of computing infrastructure and what it means for the U.S. energy sector to be ready (Fist and Datta, 2024). A Goldman Sachs analysis examined the AI capabilities and ecosystems among global AI incumbents and attempts to quantify the power impact of global data center development and AI model deployment (Singer et al., 2024; Cohen and Lee, 2023).

These studies project the future energy supply and demand of AI largely without regard to geographic or geopolitical differences. They rely particularly on recent historical data and trends in new energy generation capacity, transmission, and system performance in different countries as indicators of future capabilities. This can be measured, in part, by metrics such as the projected future capacity and transmission buildout, existing and projected excess capacity, reliability rates, curtailment rates, current and projected future demand, completion rates, and environmental targets, including local air pollution and greenhouse gas commitments.

In the United States, utilities typically develop and publish Integrated Resource Plans (IRPs) that describe their approach to meeting projected future electricity demands using both supply and demand-side resources to ensure reliable electricity amid cost, emissions, equity and other goals. Most states require utilities to create and publicly file IRPs, typically with 20-year horizons every 2 to 3 years (National Conference of State Legislatures, 2024). Such data helpfully reflects the conditions of current electricity market and the regulatory processes, and at the same time, send signals to consumers and other market stakeholders on opportunities and risks.

In the U.S., and to a lesser extent in other liberalized electricity markets such as parts of Canada, the UK, and Australia, utilities or system operators develop resource plans or capacity forecasts that inform infrastructure decisions. In the U.S., IRPs provide critical data for understanding how electricity providers intend to meet rapidly growing AI-related demand. These documents reflect market conditions and regulatory constraints, especially in light of the mixed outcomes of electricity market deregulation in recent decades (Borenstein and Bushnell, 2015; Xu et al., 2020).

However, such data may not hold the same analytical value for assessing AI energy development capabilities among key competitors such as China and the Gulf Countries that have far more centralized and authoritarian decision making. In these systems, a primarily data-driven approach could overestimate the transparency of the market conditions and underestimate the state's capacities (Boullenois et al., 2025; Seznec and Mosis, 2021). In truth, the market (and data that describes the market) does not adequately capture the financial, political, and physical resources that these countries could leverage to achieve ambitions and overcome challenges (Gunter et al., 2025). In these countries, the state has strong influence over a much broader aspect of the economy and society, and, relative to the U.S., the commercial activities are subject to much more operational controls.

As the examples in the next section show, governments in these countries allocate resources centrally for initiatives that are often pursued *against* market trends, and by *bypassing* market mechanisms to overcome key constraints (Bal and Gill, 2020).

In this Working Paper, we discuss the relative capacities for U.S. competitors to allocate resources for strategic objectives like AI supremacy, and we use historical examples to explore how state-led AI development may be achieved through non-market mechanisms and distinct institutional contexts. We therefore suggest that the approaches used to assess the potential for AI and energy competitiveness in these countries needs to be different from the approach taken in Western democracies. We conclude with two recommendations on how analysis of AI and energy competitiveness of these states should be tailored to reflect their unique political economies.

## Historical Examples of State-Led Approaches to Infrastructure Development

Non-market mechanisms have played pivotal roles in shaping infrastructure ambitions in China and Gulf Countries. These mechanisms include direct government intervention and resource allocation led by State-Owned Enterprises (SOEs)—firms controlled or funded by the government – and have enabled resource mobilization, overcoming market limitations to achieve strategic objectives. Below, we describe examples from the past 25 years that illustrate how non-market approaches in these countries have influenced infrastructure development.

While these examples illustrate how governments have used non-market mechanisms to mobilize resources, they are not intended to be comprehensive or systematically representative. We selected cases that were well-documented in English-language sources and visibly linked to national strategy. Other examples may reveal different dynamics, particularly in countries where project documentation is less transparent or less internationally visible.

*China's South-North Water Transfer Initiative*

In the 1990s China faced severe water scarcity due to a rapidly growing population and a rapidly improving standard of living that caused demand for water to far outpace supply. In response, China introduced the South-North Water Transfer Initiative, a 50-year project to transfer water from the country's water abundant south to the populous but dry northern cities (Freeman, 2011).

Officially launched in 2002, the Water Transfer Initiative demonstrates China's determination and capabilities to carry out political and economic priority projects. The three water transfer routes under this initiative covered multiple river systems, linking the Yangzi, Yellow, Haihe and other major river systems, and overcoming natural barriers with substantial investment commitments and colossal engineering designs. The Eastern Route leveraged existing canals and waterways and aimed to improve water distribution infrastructure over the distance of 1,200 kilometers. The Eastern and Central Routes also both run under the Yellow River, the $6^{th}$ longest river in the world. The Western Route, perhaps the most ambitious effort, aims to move water over hundreds of kilometers through the Bayankala Mountain to replenish the Yellow River for irrigation and economic development in western China (Freeman, 2011).

The scale of China's Water Transfer Initiative is not merely shown in the 44.8 billion cubic meters of water expected to be transferred every year, or the US $ 62 billion estimated project cost. It is further reflected in China's commitment to a 50-year ambition and a willingness to mobilize construction, financial and administrative resources from across the country to implement a goal that persists despite administration changes, environmental concerns, and local opposition (Freeman, 2011).

Unlike a market-based response that would rely on price signals or local investment to drive water distribution, the South-North Water Transfer Initiative was pursued as a centrally mandated national priority despite its limited economic return, enormous public cost, and long investment horizon. The project relied on direct government planning and funding and proceeded despite environmental opposition, interjurisdictional coordination challenges, and a lack of evidence that northern cities or industries could (or would) pay for the full cost of water delivery (Freeman, 2011).

*China's West-East Energy Transfer Initiative*

The regional imbalance of natural resources also limits development of China's energy sector. The Chinese government replicated the above whole-of-nation approach in the West-East Electricity Transmission Initiative at the turn of the $21^{st}$ century (Gibson, undated). Arguably motivated by the experience in water diversion projects, China launched the cross-country power transmission plans in the $10^{th}$ Five-Year Plan (2000-2005), bringing electricity from the resource-rich western provinces to its economically vibrant eastern coast. This energy transmission initiative involved boosting generation capacity from coal and hydro facilities and constructing three major electricity transmission corridors to serve North China, the Yangtze River Delta and the Pearl River Delta respectively. Each of these corridors were designed to carry over 40 GW by 2020 and meet 40% of China's electricity needs concentrated in seven recipient provinces. In executing the Electricity Transmission Initiative, China simultaneously established itself as the leader in the Ultra-High Voltage Direct Current (UHVDC) technology — a method of transmitting electricity over very long distances with minimal loss. The longest UHVDC line in the world was constructed by China to connect dams from upstream Yangtze River directly to the river's pacific outlet in Shanghai, a 1,287 mile transmission line with capacity to transfer 6.4 GW of electricity (Gibson, undated).

Since the announcement of the Carbon Neutrality 2060 Goal in 2020, China has also implemented a range of renewable energy projects under the umbrella of the power transmission initiative (McGrath, 2020). These include Coal-to-Solar transition in provinces around the Gobi Desert, West-to-East Gas Transmission Project (4,000 km gas pipeline originating from the Xinjiang region), and several eastward Green Hydrogen Transmission Lines ("Sinopec to Build First "West to East" Green Hydrogen Transmission Line in China," 2023; "Coal-to-green transition redefining China's west-to-east power transmission," 2024) . These projects were frequently spearheaded by major SOEs in alignment with national strategies, achieving the construction-to-operation cycle at an impressive pace despite their substantial scale.

For example, the 1,287-mile Jinping–Sunan UHVDC transmission line—from hydro dams in western Sichuan to eastern Jiangsu—was completed in under three years (2009–2012), with a capacity of 7.2 GW. This rapid timeline and scale are characteristic of China's broader West-East Electricity Transmission Initiative, which aimed to build multiple 40+ GW corridors within two decades (Paulson Institute, 2015).

 By contrast, the U.S.'s TransWest Express Transmission Project, a 732-mile high-voltage transmission line with a 3GW capacity meant to deliver wind power from Wyoming to the Southwest, took over 15 years simply from proposal to groundbreaking due to lengthy permitting, environmental reviews, and inter-jurisdictional coordination (U.S. Department of Energy, undated). Construction only began in 2023, despite the project's critical role in renewable energy integration.

China's West-East Transfer Initiative bypassed typical market allocation by directing investment to remote western provinces that had abundant generation potential but limited local demand or commercial viability. Rather than allowing market prices or utility profits to determine infrastructure expansion, the central government mandated the buildout of long-distance ultra-high voltage (UHVDC) transmission corridors, with costs borne by SOEs and justified by national development goals. The state essentially created demand by compelling eastern provinces to absorb western-generated electricity, regardless of regional price competitiveness or local utility interest.

*Saudi Arabia's Neom Mega-City*

Similarly, Saudi Arabia poured vast financial and government resources behind the futuristic *Neom* mega-city project as part of Crown Prince Mohammed bin Salman's Vision 2030. Since its announcement in 2017, Neom was planned to serve as a proud example of a modernity and competitiveness, featuring a linear city stretching over 170 kilometers with no traditional roads or cars. The Line – thanks to the narrow and futuristic layout – would be approximately the length of Belgium upon completion. The financial commitment to achieve this vision has ballooned to US $1 trillion, despite the scaled-back project scope (Magid et al., 2024).

The project exemplifies a state-driven approach, aiming to diversify Saudi Arabia's economy away from oil dependency and have more dynamic long-term development in Saudi Arabia (Magid et al.,2024). There are concerns about the feasibility and enormous costs involved, especially when the site's geography and climate does not naturally support such an extensive endeavor, and it requires displacement of original tribal communities. Against the market headwinds of low oil prices and challenges to securing foreign investment, the Saudi government and its sovereign wealth fund, the Public Investment Fund, have reiterated their support for Neom, including the proposal to list the Neom in an initial public offering (Saba, 2024).

In sum, Neom illustrates a clear case of top-down investment detached from normal market logic. The city's geographic location, scale, and projected infrastructure costs are poorly aligned with current population density, private investor appetite, or natural resource constraints. The project proceeds largely through state and sovereign wealth fund financing, not because there is proven private-sector demand or economic return, but because it fulfills a strategic vision set by the Crown Prince. In fact, private investment has lagged, and several components have been scaled back due to cost overruns—yet state investment continues, showcasing the use of public capital to override market signals.

## Implications of State-Led Approaches to Future AI Energy Demands

These distinctions of market-government roles in the U.S. and competitor countries carry important implications for assessing the capabilities and policies for meeting future energy demands of AI. Researchers and policy professionals in democratic societies may naturally turn to market signals, but may underestimate the extent to which resources could be rapidly coordinated and mobilized by the government despite these signals.

AI's energy supply and demand in competitor countries will likely face similar non-market influences. Artificial Intelligence is recognized as a strategic area of economic development and national competitiveness in China and the Gulf Countries. It is therefore reasonable to take into consideration these ambitions not only considering the current policy efforts, but also the non-market levers available to governments for achieving these goals in these countries.

In Saudi Arabia, the Vision 2030 strategy already led to the creation of the Saudi Data and AI Authority in 2019 (Gebauer and Smith, 2023). The Authority holds the explicit ambition to "position Saudi Arabia as the global hub where the best of data and AI is made reality" (Saudi Data and AI Authority, 2025). In 2023, the Saudi Crown Prince Mohammad bin Salman clarified the pathway forward with a US $200 million fund for early investments in local and international technology start-ups, including Artificial Intelligence ("Saudi Arabia launches $200m fund for early investment in high-tech companies", 2023). In November 2024, Saudi Arabia launched a state initiative "Project Transcendence" with a US $100bn financial promise, seeking to bring Saudi Arabia to the AI forefront among regional competitors (Newman et al., 2024). Against this backdrop, the partnership by the Saudi Public Investment Fund and Google to build an AI hub may prove to be the first of many such ventures (de Chant, 2024).

Big AI spending in the Gulf states is not limited in Saudi Arabia. For instance, the United Arab Emirates (UAE) announced a US $100bn fund, MGX, in partnership with BlackRock and Microsoft. This fund will follow the UAE government's first-in-the-region National Artificial Intelligence Strategy to improve competitiveness in the global race for AI leadership. The government-funded Technology Innovation Institute was behind the development of Falcon, which is in competition with American and French large language models in the AI race (United Arab Emirates Minister of State for Artificial Intelligence, 2025; "Abu Dhabi throws a surprise challenger into the AI race," 2023). In parallel, UAE has poured resources to build an AI coalition that involves heavyweights in the technology landscape (G42, 2025).

The most conspicuous example is G42, technology holding company formed in 2018 to implement the national strategy (G42, updated). Sheikh Tahnoon bin Zayed Al Nahyan, the national security adviser and an influential member of Abu Dhabi's royal family, is steering the ship as chair of G42, which is quickly building partnerships with big tech companies on top of securing support from the Emirati sovereign wealth fund Mubadala. In April 2024, Microsoft announced a US $1.5bn investment in G42, and this significant partnership could help accelerate the UAE's AI agenda with a more dynamic technology ecosystem (Microsoft, 2024).

The appealing position of Saudi Arabia, UAE and potentially other Gulf countries rests as much on the deep pockets of sovereign wealth funds as in the room for maneuver in energy supply. For example, under the Vision 2030 plan, Saudi Arabia aims to derive more than half of power from renewable sources. This ambition could translate into capacity expansion to the level of 130GW, with nearly 100GW of the total capacity from solar and wind energy (Alfehaid and Young, 2024). This renewable energy target is the most ambitious among Gulf Cooperation Council countries. Leveraging the multiple programs and implementation vehicles in the energy sector, Saudi Arabia could rapidly deploy resources into energy infrastructure projects to support *hyperscalers*—massive technology companies that operate large-scale cloud and AI infrastructure—and anchor a local AI ecosystem.

The Chinese government released an AI focused national development plan almost at the same time as the UAE. In 2017, China's State Council issued the Next Generation Artificial Intelligence Development Plan, officially considering Artificial Intelligence as a strategic priority (State Council of the People's Republic of China, 2017). China is already a leader in the field of Artificial Intelligence, ranked 2$^{nd}$ in terms of the number of patents and journal publications, and the Development Plan charted out targets for 2020, 2025 and 2030 in various levels of AI capability and application development. This plan also made clear that China will continue to leverage government fiscal resources to support and scale up early-stage AI projects, and develop national-level AI labs to complement ventures in the private sector. In the same year, the Ministry of Technology has established a Next Generation AI Development Task Force, coordinating resources from 15 ministries and facilitate the implementation of key AI projects. Additionally, China convened 27 academic and business sector AI experts to form the Next Generation AI Strategy Consultation Committee, steering political leaders and ministerial stakeholders on the course of capability development (Institute of Automation at the Chinese Academy of Sciences, 2017; "Report Release: 2019 Report on the Development of the Next Generation Artificial Intelligence in China," 2019).

Another way to validate China's policy intention in Artificial Intelligence – the senior leadership of the Chinese Communist Party had several announcements to make clear the strategic importance of Artificial Intelligence for China's technological transformation. For instance, the Party Politburo – top leadership body of the Chinese Community Party – infrequently convene its members for group training sessions on important policy or political issues, and in October 2018, Xi Jinping chaired the 9$^{th}$ Session focusing on the current development and trends for AI. There are only 24 top political leaders in the Politburo, including Xi Jinping, who used this session to reiterate AI's importance as a national strategy ("Xi Jinping Chairs the 9th Collective Study of the Political Bureau of the CPC Central Committee," 2018). This policy message has been consistent since 2018, and can be seen in Xi Jinping's public messages at the 2024 World AI Expo ("Xi Jinping Delivers the Message of Congratulation to the 2024 World Intelligence Expo," 2024).

China has already launched AI analogues to its water and energy initiatives. One of the most significant AI initiatives implemented so far is the East Data West Computing Initiative. Launched in 2022 by China's central planning agency- National Development and Reform Commission, this computing-infrastructure-focused, national initiative is elevated to the status of the aforementioned national initiatives for water transfer and energy transmission ("Three Questions on the East-Data-West-Compute Initiate," 2022). China planned eight national-level computing clusters, each comprised of a system of computing infrastructure and supporting technology investment. This means China aims to divert more and more computing tasks from the economic activities on the east coast to its western provinces, where energy and land are much less constrained. The planning commission explicitly stated that China can "leverage its advantage in national-scale planning to optimize resource allocation", and that computing infrastructure will be the central economic infrastructure of the future, just as water resources and electricity in the agricultural and industrial eras. As of the first half of 2024, China has directly invested over US $6.1 billion in the eight national computing clusters, and attracted over US $28 billion in additional funding for these state-of-the-art facilities ("Direct Investment

Exceeded RMB 43.5bn in the Eight National-Level Compute Clusters of the East-Data-West-Compute Initiate," 2024). Government statistics from China showed the latency between computing cluster nodes in West and East China to be under 20 milli-seconds, and the Power Usage Effectiveness (PUE) —a standard metric for energy efficiency in data centers, where a value closer to 1.0 means less energy is wasted on cooling and overhead—for newly built data centers achieved as low as 1.04 (For reference, Google's reported PUE average is 1.10 (Google Data Centers, 2025)).

As the East Data West Compute Initiative goes into implementation, we start to see footprints of local governments and SOEs. In the Southwest Compute Hub, the Chongqing municipality is heavily involved in the planning and resource coordination, with JD Group acting as the key private sector development partner ("Chongqing banks on East Data West Computing for new impetus," 2022). The State-Owned Assets Supervision and Administration Commission of the State Council also reported rapid response from the three telecom operators in 2022, showing the oligopolist telecom SOEs' involvement in the national initiative from an early stage (Kenji, 2022; "Telecom Operators React Vigorously to Nation's East-West Plan," 2022).

## Recommendations for Understanding and Advancing U.S. Competitiveness

The difference in capacities for resource allocation between the U.S. and its competitors leads to two recommendations: first, that analysts use geopolitically informed data and analysis to understand future competitiveness; second, that policymakers consider both market and non-market options to increase U.S. competitiveness in the face of those differences.

*Use Geopolitically Informed Data and Analysis to Understand Competitiveness*

The relationship between AI energy demand and supply will vary significantly by region and be shaped by the governance structures and economic models of individual countries. Using data and methods that are sensitive to these differences is critical for researchers aiming to analyze global trends in AI infrastructure and energy supply.

In the U.S., private technology companies, particularly those in the data center and cloud computing industries, lead the development of AI-related infrastructure. These firms invest in advanced energy-efficient technologies, renewable energy sources, and partnerships with utilities to address the growing demand for compute power. This suggests the following key sources of information:

- Public disclosures from tech giants such as Amazon, Google, and Microsoft regarding energy usage and sustainability goals.
- Industry reports on private investments in AI-specific hardware and data centers.
- Utility system performance data and future plans for renewable energy integration and grid modernization, often outlined in regulatory filings and IRPs.

In China, AI and energy initiatives are tightly woven into national policy frameworks, such as the Five-Year Plans and sector-specific strategies like the Next Generation AI Development Plan. These documents establish top-level priorities, which are implemented at regional and agency levels through localized plans and projects. This suggests a different set of data sources, including:

- The Five-Year Plan, which outlines macro-level goals for AI infrastructure and renewable energy development, often emphasizing AI as a tool for economic growth and global competitiveness.
- Regional government and ministry policy action projects that translate these high-level strategies, such as constructing hyperscale data centers or upgrading power grids to support AI compute demands.
- SOE project reports, such as those from China State Grid and key financial institutions like the China Development Bank act as primary executors, funneling resources into prioritized sectors.

In the Gulf States, AI competitiveness is framed within broader national visions, such as Saudi Arabia's Vision 2030, and spearheaded by sovereign wealth funds like Saudi Arabia's Public Investment Fund, the Abu Dhabi Investment Authority, and Qatar Investment Authority. These funds finance not only AI research and startups but also the energy infrastructure necessary to support large-scale compute operations. This suggests a different set of data sources:

- Sovereign wealth fund investment strategies, national economic visions, and energy project disclosures.
- Progress reports of infrastructure projects tied to sovereign wealth fund investments in the Gulf States.

To accurately assess the interplay between energy and AI, researchers must also interpret these data with a nuanced understanding of the pressures and incentives shaped by regional political and economic systems. In the U.S. and other western democracies, reported figures on energy capacity and AI compute loads are typically market-driven and subject to independent verification, reflecting more private-sector accountability. In countries like China and the Gulf States, where centralized or autocratic governance predominates, the reported figures and plans often serve political and strategic purposes as well, and may lack third-party validation, making them less directly comparable without adjustment.

In particular, in centralized systems, policy announcements, development goals, and capacity reports are often crafted to signal ambition, compliance, and alignment with national strategies, rather than reality. In China, reports tied to AI and energy development—such as those from local governments or SOEs—may overstate readiness or capacity to meet Five-Year Plan objectives (Davey, 2025). Reported capacities and milestones in these regions may be influenced by incentives to project alignment with central policies and to demonstrate rapid progress, which can result in data that prioritizes appearances over operational feasibility (Zhu, 2019). In the Gulf

States, publicized metrics on AI investments or energy infrastructure often aim to showcase national modernization and competitiveness, serving both internal and external audiences.

Therefore, researchers and policymakers need to interpret and account for the performative nature of data in centralized economies, where policy announcements and development figures often aim to reinforce internal narratives of progress or external projections of strength. Their interpretations will necessarily be limited by the opacity of state-led systems and the performative nature of published figures. Without independent verification, even sophisticated analysis must be cautious, relying on qualitative and subjective interpretation of state intent, which can introduce uncertainty and potential bias into assessments of actual capacity or progress.

*Assess Market and Non-Market Options to Increase U.S. Competitiveness*

The speed and scale at which China and the Gulf States might deploy energy and AI megaprojects has led to a growing frenzy of activity in the U.S. public and private sector. Leaders in AI firms have approached the U.S. government to find paths to expedite efforts to modernize the grid and generate new power (Boak, 2024), leading to the creation of the "White House Roundtable on U.S. Leadership in AI Infrastructure" (White House, 2024). Others have gone further to suggest that the U.S. should support the rapid development of datacenters adjacent to power plants to circumvent lengthy wait times to connect to the grid ( "US AI task force co-chair asks FERC to support co-located data centers – letter", 2024). In the ongoing efforts to gain an energy edge against competitors, AI giants in the U.S. are sourcing power from previously unthinkable places. Microsoft's desire for stable and clean power supply for its data centers and AI ambitions led to the reopening of the Three Mile Island nuclear plant, despite the infamous reactor meltdown incident (Mandler, 2024). Their expansion of AI data centers has also undermined their climate commitments. Alphabet, Google's parent company, reports for example that due to AI, its operations have not been carbon neutral since 2023 (Rathi, 2024).

These approaches all speak to a common question: How can the more market-based U.S. energy sector compete with centralized governments in supplying the energy demands of AI? Some have suggested that the urgency of the AI race, and the view of AI as a utility itself (Metz and Mickle, 2024), merits a shift towards more centralized decision making and non-market mechanisms in the U.S. One potential strategy is for the federal government to create enabling conditions through administrative green lights and targeted investment incentives. This approach allows market innovators to thrive while drawing inspiration from the sovereign investment funds of Gulf Cooperation Council countries. Alternatively, forming consortia to lead projects can send strong market signals, as seen in Saudi Arabia and the UAE, where governments coordinate AI ecosystem conditions. In the U.S., working groups that already touch on the AI and Energy issues could expand their roles beyond being vehicles of advice and recommendation, and start actively building shared infrastructure. Government participation in such collective implementation efforts helps address economic incentive barriers upfront, without going to the extreme of creating outright SOEs. The government could also take a more direct role in power infrastructure, akin to historical precedents in utilities. Although this approach promises to pull more potential levers at the disposal of the U.S. government, it is likely that such interventions need to be justified by the widespread benefits of Artificial Intelligence, and would impose certain guidelines for AI development and use.

While the mission to accelerate new energy generation may benefit from the speed and power of centralized decision making, this is not necessarily the U.S.'s strength and takes a narrow approach to the AI-energy nexus. The broader question is, what lengths the U.S. should go to and what levers should it pull to win the AI-energy competition, amid all the competing demands and goals for electricity (Secretary of Energy Advisory Board, 2024), including reliability, equity, and sustainability?

Here, the existing U.S. approach may have strengths. The IRP process - which competing nations often do not use - is intended to find least cost ways to meet energy demand while also meeting a myriad of goals and with a range of supply-side, demand-side, and operational methods. Indeed, many argue that indeed the U.S. may be better able to meet the AI demands of electricity by lowering demands through innovation in materials, processes, and algorithms (Hutson, 2024; Argerich and Patiño-Martínez, 2024).

However, incorporating AI energy demands into usual market-responsive processes may pose several challenges. For one, there may be a lag of many years between an IRP decision and a new power resource coming online. In contrast, data centers construction timeline can be 3 years (and sometimes much less) from groundbreaking to operation, much faster than the typical wait for utility power (Obando, 2023).

Utilities have also historically developed sophisticated tools for managing and reducing consumer demand, including time-of-use pricing and demand response programs (U.S. Department of Energy, undated). AI datacenters instead require constant, high-intensity power loads that don't easily conform to conventional demand management strategies (Fist and Datta, 2024). Unlike residential or commercial users who can shift their energy consumption to off-peak hours, AI computational workloads often demand consistent power supply to maintain performance. The sheer scale of these power demands often exceeds that of traditional industrial users, and the time-sensitive nature of AI computations means that standard price signals may have limited effectiveness in modifying usage patterns (Pyle, 2025).

And, while utilities have traditionally influenced consumer behavior through rate design, they find themselves with limited leverage to drive technical innovation in AI hardware efficiency. The major tech firms best positioned to improve AI energy efficiency may be relatively insensitive to price signals, given the high value and competitive importance of AI computations (O'Donnell, 2025a). This creates a disconnect between utility planning and the rapid pace of AI development and deployment.

This misalignment suggests the need for novel approaches to both demand management and innovation incentives. The utility sector may need to develop specialized rate structures and demand management programs specifically tailored to AI workloads. New mechanisms might be required to foster collaboration between utilities and AI firms on efficiency improvements, potentially including frameworks for sharing the benefits of innovation between energy providers and technology developers. The fundamental challenge lies in creating effective feedback loops between energy costs and AI innovation in a context where traditional utility tools may prove insufficient.

Finally, conventional planning methods may be ill-suited to handle the deep uncertainty of rapidly evolving domains like the AI-energy nexus (Lempert et al., 2003). Unpredictable variables—such as technological breakthroughs, geopolitical developments, and shifting societal priorities—can significantly alter trajectories in ways that are difficult to foresee. For example, the release of highly efficient large language models in early 2025 (e.g., DeepSeek-V2) suggests that future advances in AI may not depend solely on exponentially increasing compute or energy use. Emerging trends—such as sparse architectures and other efficiency-enhancing techniques—indicate that AI capabilities could, at least in part, become decoupled from energy consumption. However, energy efficiency does not necessarily mean lower total energy use: it means less energy is required per unit of capability. If the demand for AI capability continues to grow—as seems likely—overall energy consumption would still increase substantially. Meanwhile, training and deploying frontier models will continue to require massive infrastructure, particularly for foundational model development and large-scale inference (O'Donnell, 2025b).

To navigate this uncertainty, analysts should consider using methods of decision making under deep uncertainty (DMDU) to explore how different assumptions about deeply uncertain future characteristics affect outcomes, to identify robust and adaptive policies, and to communicate uncertainties to stakeholders, helping them understand the trade-offs and risks associated with different policy options (Marchau et al., 2019).

## Conclusion

The AI-energy nexus presents a transformative, yet complex challenge shaped by diverse regional political economies. Meeting this challenge requires asking broad questions about how the energy demands of AI should be met and answering those questions in the context of each region's unique governance structures, priorities, and institutional constraints.

Recent breakthroughs in model efficiency, multi-modality, and open-source capabilities may accelerate AI deployment beyond what current infrastructure forecasts assume. These advances raise important questions about the evolving relationship between energy, compute, and strategic advantage. Future competitiveness may depend not just on megaprojects, but on how quickly countries can adapt to shifting technological paradigms.

Ultimately, however, the race to power the most advanced AI is a means to other ends. Like electricity or roads or the Internet – public goods to which AI is often compared – the value is not in having it but in using it to enable other goals that have long stymied us – social equity, economic growth, innovation, environmental sustainability, health and wellbeing, and safety and security.

## Abbreviations

| | |
|---|---|
| AI | Artificial Intelligence |
| DMDU | Decision Making Under Deep Uncertainty |
| AGI | Artificial General Intelligence |
| GHG | Greenhouse gas |
| GW | Gigawatt |
| IRP | Integrated Resource Plans |
| Km | Kilometer |
| RDM | Robust Decision Making |
| RMB | Ren Min Bi (Chinese Yuan) |
| SOE | State-Owned Enterprises |
| PUE | Power Usage Effectiveness |
| UHVDC | Ultra-High Voltage Direct Current |
| UAE | United Arab Emirates |
| UK | United Kingdom |
| U.S. | United States |

# References

"Abu Dhabi throws a surprise challenger into the AI race," The Economist, September 21, 2023. As of April 15, 2025: https://www.economist.com/business/2023/09/21/abu-dhabi-throws-a-surprise-challenger-into-the-ai-race.

Aldrete, Bella, Jacob Ward, and Emily Pandise. "The AI industry is pushing a nuclear power revival — partly to fuel itself," *NBC News*, March 7, 2024. As of December 17, 2024: https://www.nbcnews.com/tech/tech-news/nuclear-power-oklo-sam-altman-ai-energy-rcna139094

Alfehaid, Rawan M., and Karen E. Young. "Saudi Arabia's Renewable Energy Initiatives and Their Geopolitical Implications," Center on Global Energy Policy, SIPA, Columbia University, October 29, 2024.

Argerich, Mauricio Fadel, and Marta Patiño-Martínez. "Measuring and Improving the Energy Efficiency of Large Language Models Inference." IEEE Access, 2024. As of December 18, 2024: https://ieeexplore.ieee.org/abstract/document/10549890

Bal, Ravtosh, and Indermit S. Gill. "Policy approaches to Artificial Intelligence based technologies in China, European Union and the United States." Center on Global Energy Policy, SIPA, Columbia University, 2020

Boak, Josh. "Top AI business leaders meet with Biden administration to discuss the emerging industry's needs," *Associated Press*, September 12, 2024. As of December 17, 2024: https://apnews.com/article/biden-white-house-ai-google-openai-nvidia-33954185e8afb49a13a86d8fa4b9b14b

Boullenois, Camille, Kratz, Agatha, and Daniel H. Rosen. "Far From Normal: An Augmented Assessment of China's State Support," Rhodium Group, March 17, 2025.

Bourzac, Katherine. "Fixing AI's energy crisis." *Nature*, 2024. As of December 18, 2024: https://www.nature.com/articles/d41586-024-03408-z

Cheng, Yuk-shing, Man-kit Chung, and Kam-pui Tsang. "Electricity Market Reforms for Energy Transition: Lessons from China." Energies 16, no. 2, 2023: 905.

"China is building nuclear reactors faster than any other country," The Economist, November 30, 2023. As of December 19, 2024: https://www.economist.com/china/2023/11/30/china-is-building-nuclear-reactors-faster-than-any-other-country

"Chongqing banks on East Data West Computing for new impetus," CGTN, October 5, 2022. As of December 19, 2024: https://news.cgtn.com/news/2022-10-05/Chongqing-banks-on-East-Data-West-Computing-for-new-impetus-1dRDQStm9kA/index.html

"Coal-to-green transition redefining China's west-to-east power transmission," Xinhua News, September 14, 2024. As of December 19, 2024: https://english.news.cn/20240914/8de326df8041428187bc218596316f73/c.html

Cohen, Jared, and George Lee. "The generative world order: AI, geopolitics, and power." Goldman Sachs. December 14, 2023. As of December 19, 2024: https://www.goldmansachs.com/insights/articles/the-generative-world-order-ai-geopolitics-and-power

Davey, Alexander. "Read the CCP's policy priorities: A glimpse into the black box of China's policymaking process," Mercator Institute for China Studies, March 5, 2025. As of April 14, 2025: https://merics.org/en/report/read-ccps-policy-priorities-glimpse-black-box-chinas-policymaking-process.

de Chant, Tim, "Google is opening an AI hub in oil-rich Saudi Arabia," *TechCrunch*, November 5, 2024.

"Direct Investment Exceeded 43.5bn in the Eight National-Level Compute Clusters of the East-Data-West-Compute Initiate ["东数西算"八大国家枢纽节点直接投资超过435亿元]," Xinhua News, August 29, 2024. As of December 19, 2024: http://news.cn/fortune/20240829/5b294ff0e0dd4f24b33fbf092896c280/c.html

Fist, Tim and Arnab Datta, "How to Build the Future of AI in the United States", *Institute for Progress,* October 23rd 2024. As of December 19, 2024: https://ifp.org/future-of-ai-compute/#challenges-to-building-in-america

Freeman, Carla, "Quenching the Thirsty Dragon: The South-North Water Transfer Project—Old Plumbing for New China?" Wilson Center. July 7, 2011. As of December 19, 2024: https://www.wilsoncenter.org/publication/quenching-the-thirsty-dragon-the-south-north-water-transfer-project-old-plumbing-for-new

Hutson, Matthew. "1-bit LLMs Could Solve AI's Energy Demands," IEEE Spectrum, May 30, 2024. As of December 18, 2024: https://spectrum.ieee.org/1-bit-llm

Kenji Kawase, "Chinese telecom groups shift focus to 'east-west' data project," *Nikkei Asia*, March 30, 2022. As of December 19, 2024: https://asia.nikkei.com/Business/Telecommunication/Chinese-telecom-groups-shift-focus-to-east-west-data-project

G42, "About G42," webpage, last updated April 15, 2025. As of May 05, 2024: https://www.g42.ai/about

Gibson, David Tyler. "Interactive: China's West-East Electricity Transfer Project," Wilson Center, undated. https://www.wilsoncenter.org/publication/interactive-chinas-west-east-electricity-transfer-project

Gebauer, Sarah L., and Gregory Smith. "From Soccer to AI, Saudi Arabia Spends to Win." RAND Corporation, 2023. As of December 19, 2024:https://www.rand.org/pubs/commentary/2023/12/from-soccer-to-ai-saudi-arabia-spends-to-win.html

Google Data Centers, "Efficiency," webpage, last updated April 15, 2025. As of May 05, 2024: https://www.google.com/about/datacenters/efficiency/

Gunter, Jacob, Brown, Alexander, Chimits, François, Hmaidi, Antonia, Vasselier, Abigaël, and Max J. Zenglein. "Beyond overcapacity: Chinese-style modernization and the clash of economic models," Mercator Institute for China Studies, Apr 01, 2025. As of April 14, 2025: https://merics.org/en/report/beyond-overcapacity-chinese-style-modernization-and-clash-economic-models.

Institute of Automation at the Chinese Academy of Sciences [中国科学院自动化研究所], *Strategic Advisory Committee on Next Generation Artificial Intelligence Formally Established [新一代人工智能战略咨询委员会成立]*, November 12, 2017. As of December 19, 2024: https://ia.cas.cn/xwzx/jryw/201711/t20171121_4896938.html

Lempert, R. J., S. W. Popper, and S. C. Bankes. "Shaping the Next 100 Years: New Methods for Quantitative." *Long-Term Policy Analysis, Santa Monica: RAND Corporation,* 2003.

Magid, Pesha, Hadeel Al Sayegh and Federico Maccioni. "Exclusive: Saudi Arabia prioritizes sports for Neom plans as costs balloon, sources say," *Reuters*, November 14, 2023. As of December 19, 2024: https://www.reuters.com/world/middle-east/saudi-arabia-prioritizes-sports-neom-plans-costs-balloon-sources-say-2024-11-13/

Mandler, C., "Three Mile Island nuclear plant will reopen to power Microsoft data centers," *NPR*, September 20, 2024. As of December 19, 2024: https://www.npr.org/2024/09/20/nx-s1-5120581/three-mile-island-nuclear-power-plant-microsoft-ai

Marchau, Vincent AWJ, Warren E. Walker, Pieter JTM Bloemen, and Steven W. Popper. "Decision making under deep uncertainty: from theory to practice." *Springer Nature*, 2019.

Maslej, Nestor, Loredana Fattorini, Raymond Perrault, Vanessa Parli, Anka Reuel, Erik Brynjolfsson, John Etchemendy, Katrina Ligett, Terah Lyons, James Manyika, Juan Carlos Niebles, Yoav Shoham, Russell Wald, and Jack Clark, "The AI Index 2024 Annual Report," AI Index Steering Committee, Institute for Human-Centered AI, Stanford University, Stanford, CA, April 2024.

McGrath, Matt. "Climate change: China aims for 'carbon neutrality by 2060'," *BBC*, September 20, 2022. As of December 19, 2024: https://www.bbc.com/news/science-environment-54256826

Metz, Cade and Tripp Mickle. "Behind OpenAI's Audacious Plan to Make A.I. Flow Like Electricity", *New York Times*, September 25, 2024. As of December 17, 2024: https://www.nytimes.com/2024/09/25/business/openai-plan-electricity.html

Microsoft, "Microsoft invests $1.5 billion in Abu Dhabi's G42 to accelerate AI development and global expansion," April 15, 2024. As of April 13, 2025: https://news.microsoft.com/2024/04/15/microsoft-invests-1-5-billion-in-abu-dhabis-g42-to-accelerate-ai-development-and-global-expansion/?msockid=3ad4dd69c0436ca821d7c86ec1046d46.

National Conference of State Legislatures, Integrated Resource Planning: A Primer for State Legislators. October 25, 2024. As of December 19, 2024: https://www.ncsl.org/energy/integrated-resource-planning-a-primer-for-state-legislators

Newman, Marissa, Love, Julia, Bergen, Mark and Christine Burke. "Saudis Plan $100 Billion AI Powerhouse to Rival UAE Tech Hub," *Bloomberg*, November 7, 2024.

Obando, Sebastian. "Data center building boom shows cracks," ConstructionDrive.com, November 10, 2023. As of December 20, 2024: https://www.constructiondive.com/news/psmj-survey-data-center-construction/699469/

O'Donnell, James, "AI's energy obsession just got a reality check," *MIT Technology Review*, January 28, 2025a. As of April 13, 2025: https://www.technologyreview.com/2025/01/28/1110599/ais-energy-obsession-gets-a-reality-check/

O'Donnell, James, "DeepSeek might not be such good news for energy after all," *MIT Technology Review*, January 31, 2025b. As of April 13, 2025: https://www.technologyreview.com/2025/01/31/1110776/deepseek-might-not-be-such-good-news-for-energy-after-all/

Paulson Institute, *Power Play: China's Ultra-High Voltage Technology and Global Standards*, April 2015. As of April 14, 2025: https://www.paulsoninstitute.org/wp-content/uploads/2017/01/PPS_UHV_English_R.pdf

Pyle, Thomas J. "Putting AI's Insatiable Electricity Demand in Perspective," Institute for Energy Research, January 7, 2025. As of April 13, 2025: https://www.instituteforenergyresearch.org/the-grid/putting-ais-insatiable-electricity-demand-in-perspective/

Qiu, Yueming, Nana Deng, Bo Wang, Xingchi Shen, Zhaohua Wang, Nathan Hultman, Han Shi, Jie Liu, and Yi David Wang. "Power supply disruptions deter electric vehicle adoption in cities in China." *Nature Communications* 15, no. 1 (2024): 6041.

Rathi, Akshat. "Google Is No Longer Claiming to Be Carbon Neutral," Bloomberg, July 8, 2024. As of December 19, 2024: https://www.bloomberg.com/news/articles/2024-07-08/google-is-no-longer-claiming-to-be-carbon-neutral?embedded-checkout=true

"Report Release: 2019 Report on the Development of the Next Generation Artificial Intelligence in China [《中国新一代人工智能发展报告2019》发布]", People's Daily, May 26, 2019. As of December 19, 2024: https://www.gov.cn/xinwen/2019-05/26/content_5394817.htm

Reuters. "US AI task force co-chair asks FERC to support co-located data centers – letter," December 6, 2024. As of December 17, 2024: https://www.reuters.com/business/energy/us-ai-task-force-co-chair-asks-ferc-support-co-located-data-centers-letter-2024-12-07/

Saba, Yousef. "Saudi crown prince says Neom mega-project to list in 2024," *Reuters*, July 26, 2024. As of December 19, 2024: https://www.reuters.com/world/middle-east/saudi-crown-prince-says-zero-carbon-city-neom-will-likely-be-listed-2024-2022-07-25/

"Saudi Arabia launches $200m fund for early investment in high-tech companies," Arab News, August 21, 2023. As of December 19, 2024: https://www.arabnews.com/node/2358236/business-economy

Saudi Data and AI Authority, "Our Achievements," webpage, last updated April 15, 2025. As of December 10, 2024: https://sdaia.gov.sa/en/SDAIA/about/Pages/Achievements.aspx

Secretary of Energy Advisory Board, "Recommendations on Powering Artificial Intelligence and Data Center Infrastructure," U.S. Department of Energy, July 30, 2024. As of December 18, 2024: https://www.energy.gov/sites/default/files/2024-08/Powering%20AI%20and%20Data%20Center%20Infrastructure%20Recommendations%20July%202024.pdf

Seznec, Jean-François, and Samer Mosis. "The Energy Transition in the Arab Gulf: From Vision to Reality," Atlantic Council, July 2021.

Singer, Brian, Carly Davenport, Allen Chang, Evan Tylenda, Neil Mehta, Joe Ritchie, Jerry Revich, Jacqueline Du, John Miller, Derek R. Bingham, Alberto Gandolfi, Timothy Zhao, Emma Jones, Brian Lee, Apoorva Bahadur, Ryo Harada, Grace Chen, Ati Modak, Brenda Corbett, Toshiya Hari, Varsha Venugopal, Madeline Meyer, Daniela Costa, Mark Delaney, Christian Hinderaker, and Xavier Zhang, "AI/data centers' global power surge and the sustainability impact," Goldman Sachs. April 28, 2024. As of December 19, 2024: https://www.goldmansachs.com/images/migrated/insights/pages/gs-research/gs-sustain-generational-growth-ai-data-centers-global-power-surge-and-the-sustainability-impact/sustain-data-center-redaction.pdf

Singh, Manish, "DeepSeek 'punctures' AI leaders' spending plans, and what analysts are saying", *TechCrunch,* January 27, 2025. As of April 12, 2025: https://techcrunch.com/2025/01/27/deepseek-punctures-tech-spending-plans-and-what-analysts-are-saying/

"Sinopec to Build First "West to East" Green Hydrogen Transmission Line in China," World Energy, April 13, 2023. As of December 19, 2024: https://www.world-energy.org/article/31250.html

State Council of the People's Republic of China [国务院], "Notice by the State Council on releasing the Development Guidelines for the Next Generation Artificial Intelligence [国务院关于印发新一代人工智能发展规划的通知]," State Council Notice –2017-35 [国发〔2017〕35号], July 20, 2017. As of December 19, 2024: https://www.gov.cn/zhengce/content/2017-07-20/content_5211996.htm

"Three Questions on the East-Data-West-Compute Initiate ["东数西算"三问]," Xinhua News, February 17, 2022. As of December 19, 2024: https://www.gov.cn/zhengce/2022-02/17/content_5674406.htm

"Telecom Operators React Vigorously to Nation's East-West Plan," China Daily. March 21, 2022. As of December 19, 2024: http://en.sasac.gov.cn/2022/03/21/c_8768.htm

United Arab Emirates Minister of State for Artificial Intelligence, "UAE National Strategy for Artificial Intelligence 2031," webpage, last updated April 15, 2025. As of May 05, 2024: https://ai.gov.ae/strategy/

U.S. Department of Energy, "Demand Response and Time-Variable Pricing Programs," webpage, undated. As of December 18, 2024: https://www.energy.gov/femp/demand-response-and-time-variable-pricing-programs

U.S. Department of Energy, "TransWest Express Transmission Project," updated. As of April 30, 2025: https://www.energy.gov/oe/articles/transwest-express-transmission-project

White House, "Readout of White House Roundtable on U.S. Leadership in AI Infrastructure," September 12, 2024. As of December 17, 2024: https://www.whitehouse.gov/briefing-room/statements-releases/2024/09/12/readout-of-white-house-roundtable-on-u-s-leadership-in-ai-infrastructure/

"Xi Jinping Chairs the 9th Collective Study of the Political Bureau of the CPC Central Committee [习近平主持中共中央政治局第九次集体学习并讲话]," Xinhua News, October 31, 2018. As of December 19, 2024: https://www.gov.cn/xinwen/2018-10/31/content_5336251.htm

"Xi Jinping Delivers the Message of Congratulation to the 2024 World Intelligence Expo [习近平向2024世界智能产业博览会致贺信]," Xinhua News, June 20, 2024. As of December 19, 2024: https://www.gov.cn/yaowen/liebiao/202406/content_6958352.htm

Xu, Chengwei, Fushuan Wen, and Ivo Palu. "Electricity market regulation: Global status, development trend, and prospect in China." *Energy Conversion and Economics* 1, no. 3, 2020: 151-170.

Zhu, Xufeng. "Central-Local Relations in China," in Yu, J., Guo, S. (eds) *The Palgrave Handbook of Local Governance in Contemporary China*, Palgrave Macmillan, Singapore, 2019.